# Tunable optical bistability of graphene-wrapped dielectric cylinders

K. Zhang[1], and L. Gao[1,2,*]

[1]College of Physics, Optoelectronics and Energy of Soochow University, Collaborative Innovation Center of Suzhou Nano Science and Technology, Soochow University, Suzhou 215006, China

[2]Jiangsu Key Laboratory of Thin Films, Soochow University, Suzhou 215006, China.

*leigao@suda.edu.cn

**Abstract**

We theoretically study the role of nonlinear surface plasmoms on the optical bistability of graphene-wrapped dielectric cylinders, within the framework of both full-wave scattering theory and the quasistatic limit. Typical hysteresis curves are observed in both near-field and far-field spectra. Moreover, we demonstrate that introducing the full wave theory results in another bistable behavior with a high applied electromagnetic field, suggesting a more explicit way in analyzing the unstable behavior of the graphene-wrapped dielectric cylinder. Furthermore, optical stable region and the switching threshold values are proved to be tunable by changing either the size, permittivity of the nanocylinder or the chemical potential of graphene, promising the graphene-wrapped dielectric cylinder a candidate for all-optical switching and nano-memories.

**Introduction**

Nonlinear optical effects, modifying the optical properties of a material system by the presence of light, play an important role in modern photonic functionalities, including ultrafast optical switching, optical transistors, optical modulation and so on [1]. However, governed by photon–photon interactions enabled by materials, optical nonlinearities are inherently weak. They are superlinearly field-dependent and able to be enhanced in material environments, which provide mechanisms for field enhancement. Hereby, plasmonic-enhanced-nonlinear structures, usually characterized by metal/dielectric composites, have been widely studied [2-4]. Such composites support nonlinear surface plasmonic resonances in the interface of the dielectric and metal [5, 6], resulting in strong electromagnetic field and then boost the field-dependent nonlinearity, which dramatically shortens the response time and allows nonlinear optical components to be scaled down in size [7].

In a nonlinear system, due to a self-feedback mechanism, bistable states under a given external condition can occur. Such as bistability of transmission in a Fabry-Perot interferometer filled with a nonlinear medium [8], and hysteretic reflection curve at a nonlinear interface [9]. Optical bistability (OB) is a way of controlling light with light [10, 11], where a nonlinear optical systems shows two different values of the local field intensity for one input intensity, and for applications, exploring the Kerr nonlinear effect, nonlinear modifications of the refractive index of the material [1], is one effective way to analyze the OB [12-14], which makes it able to realize in a single device a series of functionalities, such as optical switching, logical memory, modulation and so on, with one input power [15]. And eventually, one can realize the optical computer [16].

On the other hand, graphene, as an excellent optoelectronic material [17], exhibits an intrinsic nonlinear optical response in several frequency regimes [18-20], and works based on the nonlinearity of graphene have been down to explore the potential applications both theoretically and experimentally, such as the mode-locking fiber [21], harmonic generations [22, 23], nonlinear surface plasmons (SPs) [24-26] and so on. Besides, optical bistable behavior of graphene/graphene-based structures have been widely investigated [12, 25, 27] in 1D system. For details, Ref. 12 and 27 study the OB of graphene based on the intrinsic nonlinearity of graphene, and Ref. 25 shows a SPPs-enhanced bistable behavior in graphene. Different from these works, we theoretically study the optical bistability of the graphene-wrapped cylinders, which has already been realized in experiment [28], in both near-field and far-field spectra by generalizing linear full-wave scattering theory (FWST) [29] and quasistatic limit (QL) to nonlinear theory. Comparisons are made to find the differences between these two theories. It's found that there is only one OB no matter how strong the applied field is in the framework of the quasistatic limit, instead, double OB are found from the full-wave scattering theory. Besides, when the applied external field is small, results from the two theories match quiet well. We demonstrate that the threshold values of the single and double OB are tunable either by varying the sizes and permittivities of cylinders or changing the chemical potential. These results promise the graphene-wrapped cylinders candidate for all-optical switching, which has potential applications in optical communications and computing.

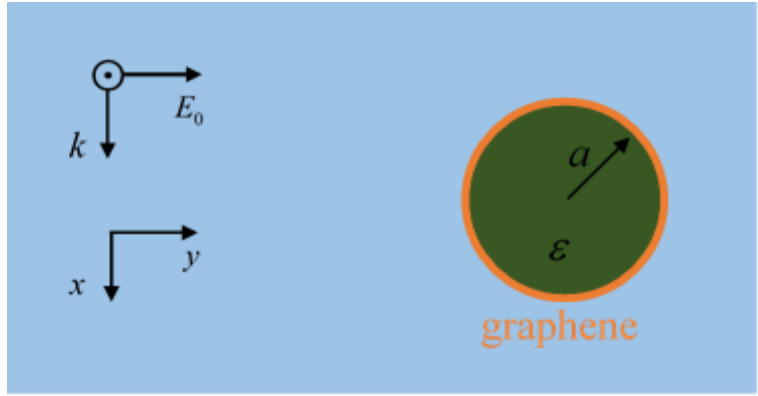

**Fig. 1. Schematic diagram of the model**

**Theoretical model and methods**

We start our work by firstly considering a linear case of a 2D system shown in Fig. 1, where a TM-polarized plane wave is applied on graphene-wrapped nano-cylinders with radius $a$ and permittivity $\varepsilon$, under the following two theories (the FWST and the QL).

**A. Linear theories for linear graphene-wrapped cylinders**

**1. Derivations based on full-wave scattering theory**

As illustrated in Fig. 1, the incident electric field is perpendicular to the *xz* plane and spreads along the *x* axis with the form of $\mathbf{E} = \hat{y} E_0 e^{ikx} \cdot e^{-i\omega t}$, where $k = k_0 \sqrt{\varepsilon_h}$ denotes the wave number in surrounding medium with permittivity $\varepsilon_h$. Based on the full wave scattering theory, the general solutions for the local electromagnetic field can be written as below [29]:

$$\mathbf{E}_i = -\frac{i}{k} \sum_{n=-\infty}^{\infty} E_n \mathbf{M}_n^{(1)}, \mathbf{H}_i = \frac{-1}{\omega\mu} \sum_{n=-\infty}^{\infty} E_n \mathbf{N}_n^{(1)}$$
$$\mathbf{E}_s = \frac{i}{k} \sum_{n=-\infty}^{\infty} A_n E_n \mathbf{M}_n^{(3)}, \mathbf{H}_s = \frac{1}{\omega\mu} \sum_{n=-\infty}^{\infty} A_n E_n \mathbf{N}_n^{(3)} \quad , \quad (1)$$
$$\mathbf{E}_c = -\frac{i}{k_1} \sum_{n=-\infty}^{\infty} F_n E_n \mathbf{M}_n^{(1)}, \mathbf{H}_c = \frac{-1}{\omega\mu_1} \sum_{n=-\infty}^{\infty} F_n E_n \mathbf{N}_n^{(1)}$$

where $E_n = E_0(-i)^n$; $\mathbf{M}_n$ and $\mathbf{N}_n$ are vector cylindrical harmonics, and the upper indices (1) and (3) represent the use of Bessel function $J_n$ and the first kind of Hankel function $H_n$. In addition, $k_1 = k_0 \sqrt{\varepsilon}$ indicates the wave number inside the cylinder, $\mathbf{E}_i, \mathbf{E}_s, \mathbf{E}_c$ together with $\mathbf{H}_i, \mathbf{H}_s, \mathbf{H}_c$ are relevant electric and magnetic incident, scattering and core fields.

Since the single layer graphene is only one-atom thick, much smaller than the particle size, the graphene coating can be theoretically characterized as a two-dimensional homogenized conducting film with surface conductivity $\sigma_g$ [30, 31], yielding the boundary conditions:

$$\hat{n} \cdot (\mathbf{E}_i + \mathbf{E}_s - \mathbf{E}_c) = 0$$
$$\hat{n} \times (\mathbf{H}_i + \mathbf{H}_s - \mathbf{H}_c) = \mathbf{J} \quad (2)$$

with $\mathbf{J} = \sigma_g \mathbf{E}_t$ being the tangential electric field induced surface current density.

Applying the boundary conditions at $r=a$, along with the component form of Eq. (1) in $\rho, \varphi, z$ directions, we can derive the coefficients as below:

$$A_n = \frac{J_n(x)J_n'(mx) - mJ_n'(x)J_n(mx) - i\sigma_g \alpha J_n'(x)J_n'(mx)}{H_n(x)J_n'(mx) - mH_n'(x)J_n(mx) - i\sigma_g \alpha H_n'(x)J_n'(mx)}$$

$$F_n = \frac{H_n(x)J_n'(x) - H_n'(x)J_n(x)}{H_n(x)J_n'(mx) - mH_n'(x)J_n(mx) - i\sigma_g \alpha H_n'(x)J_n'(mx)} \quad (3)$$

where $\alpha = \sqrt{\mu_0/\varepsilon_0\varepsilon_h}, m = \sqrt{\varepsilon/\varepsilon_h}$ and $x = ka$. Then distribution of the local electric field in the dielectric cylinder can be obtained by substituting Eq. (3) into Eq. (1). Especially, when we take $r=a$, we can achieve the local fields near the dielectric-metal interface and its form of square of modulus:

$$\mathbf{E}_{c,local} = -E_0 \sum_{n=-\infty}^{\infty} i^{n+1} F_n \left( in \frac{J(k_1 a)}{k_1 a} \hat{\mathbf{e}}_r - J_n'(k_1 a) \hat{\mathbf{e}}_\varphi \right) e^{in\varphi},$$

$$|\mathbf{E}_{c,local}|^2 = |E_0|^2 \sum_{n=-\infty}^{\infty} |F_n|^2 \left[ n^2 \left( \frac{J(k_1 a)}{k_1 a} \right)^2 + J_n'^2(k_1 a) \right] \quad (4)$$

together with the square of modulus of the tangential local field in the graphene thin layer

$$|\mathbf{E}_{lin,g}|^2 = |E_0|^2 \sum_{n=-\infty}^{\infty} |F_n|^2 J_n'^2(k_1 a). \quad (5)$$

Besides, efficiencies $Q_{sca}$ and $Q_{ext}$ for scattering and extinction can be expressed by [29]:

$$Q_{sca} = \frac{1}{x} \left( |A_0|^2 + 2\sum_{n=1}^{\infty} |A_n|^2 \right)$$

$$Q_{ext} = \frac{2}{x} \operatorname{Re} \left( A_0 + 2\sum_{n=1}^{\infty} A_n \right). \quad (6)$$

## 2. Derivations for quasistatic limit

Since the diameters of the cylinder we employed in our work are much smaller than the incident wavelength, we also put out the derivations under quasistatic limit for comparison and investigation. The electric potentials both inside

($\phi_c$) and outside ($\phi_h$) the cylinder would satisfy the Laplace equation: $\nabla^2 \phi_{c,h} = 0$, and have the general solutions:

$$\phi_c = -BE_0 r \cos\varphi$$
$$\phi_h = (-E_0 r + \frac{CE_0}{r})\cos\varphi \qquad (7)$$

To solve the coefficients $B$ and $C$, we adopt the following boundary conditions [21]:

$$\hat{n} \times [\mathbf{E}_h - \mathbf{E}_c]|_{r=a} = 0$$
$$\hat{n} \cdot [\mathbf{D}_h - \mathbf{D}_c]|_{r=a} = \rho \qquad (8)$$

where $\mathbf{E}_{c,h} = -\nabla \phi_{c,h}$ and $\mathbf{D}_{c,h} = \varepsilon_{c,h} \mathbf{E}_{c,h}$ ($\varepsilon_c = \varepsilon$) are electric field and relevant electric displacement vector inside and outside the cylinder. The symbol $\rho$ represents the surface charges, which has the relation $\rho = \nabla_s \cdot \mathbf{J}$ with the surface current density $\mathbf{J}$, and the operator $\nabla_s$ stands for the surface divergence. Combining Eq. (7) and Eq. (8), we achieve the coefficients:

$$B = \frac{2\varepsilon_h}{\varepsilon + \varepsilon_h + \gamma}$$
$$C = \frac{a^2(\varepsilon - \varepsilon_h + \gamma)}{\varepsilon + \varepsilon_h + \gamma} \qquad (9)$$

here $\gamma = i\sigma_g / (\omega a \varepsilon_0)$.

Based on $\mathbf{E}_c = -\nabla \phi_c = BE_0 (\cos\varphi \hat{\mathbf{e}}_r - \sin\varphi \hat{\mathbf{e}}_\varphi)$, we have

$$\left|\mathbf{E}_{lin,g}\right|^2_{QL} = |\mathbf{E}_c|^2 = |B|^2 |E_0|^2, \qquad (10)$$

which indicates the relation between the linear local filed in graphene and the incident field like Eq. (5).

Similarly, in small particle limit, efficiencies $Q_{sca,QL}$ and $Q_{ext,QL}$ for scattering and extinction can be expressed by [29]:

$$Q_{sca,QL} = \frac{\pi^2 (ka)^3}{4} \left|\frac{C}{a^2}\right|^2$$
$$Q_{ext,QL} = \pi(ka) \frac{\text{Im}(C)}{a^2} \qquad (11)$$

**B. Nonlinear theories for nonlinear graphene-wrapped cylinders**

Due to the dependence of local tangential field and intrinsic nonlinear property, the surface conductivity of graphene $\tilde{\sigma}_g$, within the random-phase approximation, can be written as [22, 25]:

$$\sigma_g = \sigma_0 + \sigma_3 |\mathbf{E}_t|^2, \qquad (12)$$

where $\sigma_0 = \frac{ie^2 \mu_c}{\pi \hbar^2 (\omega + i/\tau)}, \sigma_3 = -i\frac{9e^4 v_F^2}{8\pi \mu_c \hbar^2 \omega^3}$ correspond to the linear term and third order nonlinear term of the graphene conductivity with $e, \mu_c, \hbar, \tau, v_F$ being the charge of electron, chemical potential of graphene, reduced Planck constant, relaxation time and Fermi velocity. In addition, the explicit form of $|\mathbf{E}_t|^2$ in the FWST is $|\mathbf{E}_{lin,g}|^2$ described by Eq. (5) and the one under the QL is $|\mathbf{E}_{lin,g}|^2_{QL}$ in Eq. (10). By replacing the linear conductivity $\sigma_g$ in the linear derivations [i.e. Eq. (3)] by the nonlinear one $\tilde{\sigma}_g$, we have the nonlinear solutions for the nonlinear system, and we use the superscript "~" to tell the differences. On the other hand, when the nonlinear conductivity of the coated graphene is taken into account, the tangential field would definitely be nonlinear, so $|\mathbf{E}_t|^2$ would be replaced by the nonlinear one, namely, in the FWST $\tilde{\sigma}_g \approx \sigma_0 + \sigma_3 |\mathbf{E}_{non,g}|^2$,

with

$$|\mathbf{E}_{non,g}|^2 = |E_0|^2 \sum_{n=-\infty}^{\infty} |\tilde{F}_n|^2 J_n'^2(k_1 a),$$

$$\tilde{F}_n = \frac{H_n(x)J_n'(x) - H_n'(x)J_n(x)}{H_n(x)J_n'(mx) - mH_n'(x)J_n(mx) - i\tilde{\sigma}_g \alpha H_n'(x)J_n'(mx)}, \quad (13)$$

and in the QL $\tilde{\sigma}_{g,QL} \approx \sigma_g + \sigma_3 |\mathbf{E}_{non,g}|^2_{QL}$, with

$$|\mathbf{E}_{non,g}|^2_{QL} = |\tilde{B}|^2 |E_0|^2, \quad \tilde{B} = \frac{2\varepsilon_h}{\varepsilon + \varepsilon_h + i\tilde{\sigma}_g/(\omega a \varepsilon_0)}. \quad (14)$$

Meanwhile, linear scattering and extinction efficiencies [Eq. (6) and Eq. (11)] would be modified as:

$$\tilde{Q}_{sca} = \frac{1}{x}\left(|\tilde{A}_0|^2 + 2\sum_{n=1}^{\infty}|\tilde{A}_n|^2\right), \tilde{Q}_{ext} = \frac{2}{x}\mathrm{Re}\left(\tilde{A}_0 + 2\sum_{n=1}^{\infty}\tilde{A}_n\right), \quad (15)$$

in the FWST, and the one under QL is:

$$\tilde{Q}_{sca,QL} = \frac{\pi^2(ka)^3}{4}\left|\frac{\tilde{C}}{a^2}\right|^2, \tilde{Q}_{ext,QL} = \pi(ka)\frac{\mathrm{Im}(\tilde{C})}{a^2}. \quad (16)$$

**Numerical results and discussion**

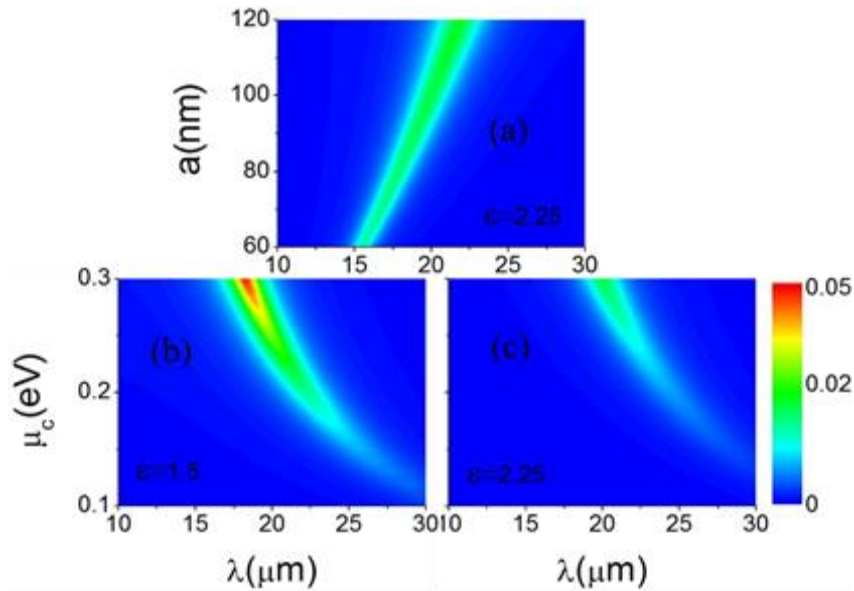

Fig. 2. Scattering efficiency for varied chemical potential, cylinder size and incident wavelength, with (a) $\varepsilon = 2.25$,

$\mu_c = 0.3eV$, **(b)** ε =1.5, *a*=100*nm*, **and (c)** ε = 2.25, *a*=100*nm*. **Other parameters are** $\varepsilon_h = 2.25$ **and** $\tau = 0.1ps$.

We are now in a position to present some numerical results. For numerical calculations, without loss of generality, we consider a graphene-wrapped dielectric cylinder embedded in pure dielectric medium with relative dielectric constant $\varepsilon_h = 2.25$, the relative permeability of the medium inside and outside the cylinder are both 1, and the carrier relaxation time is fixed at $\tau = 0.1ps$. Other parameters, such as the size and permittivity of the cylinder, as well as the chemical potential of graphene, are variable. Firstly, we investigate the linear scattering efficiency of the graphene coated dielectric cylinder based on Eq. (6) with n=1 [One should note that derivations of Eq. (11) are totally based on the Eq. (6) within small particle limit, hence we use Eq. (6) to give a general conclusion.]. In the linear calculation, we ignore the field-dependent term in Eq. (12), the surface conductivity $\sigma_g = \sigma_0$ is completely complex and its imaginary part $\text{Im}(\sigma_g) > 0$, indicating graphene a "metallic" thin layer [31], hence, plasmon resonance enhanced scattering efficiencies are observed in Fig. 2. Take a close look at Fig. 2, we conclude that with the resonant wavelength varies by the size, permittivity of cylinder and chemical potential of graphene. In more detail, for a fixed chemical potential, the resonant wavelength undergoes a red shift with the increase of cylinder sizes [see Fig. 2 (a)], accompanied by increase of resonant peak, which is mainly because the graphene layer is metal-like, the resonant wavelength tends to be size dependent. What's more, by increasing the chemical potential of graphene, which can be realized by tuning the density of the charge carriers through the external electrical gating field and/or chemical doping, one can also achieve the enhancement of scattering efficiency as Fig. 2 (b) or 2 (c) illustrated, with the chemical potential increase from 0.1eV to 0.3eV, the peak value increase together with a blue shift for the resonant wavelength, similar phenomena can be found in Abajo's work [32]. On the other hand, by comparing Fig. 2 (b) with Fig. 2 (c), we find that for fixed chemical potential, increasing the permittivity inside the cylinder will result in decrease in peak value, which shows a different phenomenon with increasing the size of cylinder or the chemical potential of graphene, however, the resonant wavelength also shows a red shift like Fig. 2 (a).

In what follows, we consider the nonlinear case with the graphene conductivity being a Kerr-like one [see Eq. (12)]. One can derive the field-dependent coefficients $\tilde{A}_n$, $\tilde{F}_n$, $\tilde{B}$ and $\tilde{C}$ by substituting Eq. (12) into Eq. (3) and Eq. (9), these coefficients will in turn feed back to the linear local field, yielding the nonlinear local field shown in Eq. (13) and Eq. (14). Hence, by fixing the input electric field intensity at $E_0 = 5 \times 10^6 (v/m)$, we firstly investigate the modulus of nonlinear local field inside graphene with $E_{non,g} = \sqrt{|E_{non,g}|^2}$, and Fig. 3 (a) depicts the field-wavelength relation. It exhibits nonlinear plasmonic resonances characterized by hysteresis curves, and the resonant peaks increase with increasing the chemical potential of graphene, which modifies the nonlinear surface conductivity [Eq. (12)] of graphene and leads to blue shift for the resonant wavelength similar to Fig. 2. (b) and Fig. 2. (c). On the other hand, the unstable region gets narrower and the threshold value decreases when the chemical potential increases from 0.3eV to 0.4eV, promising the proposed structure may realize a nonlinear nanoswitch device, whose switching frequency is tunable via varying the density of the charge carriers.

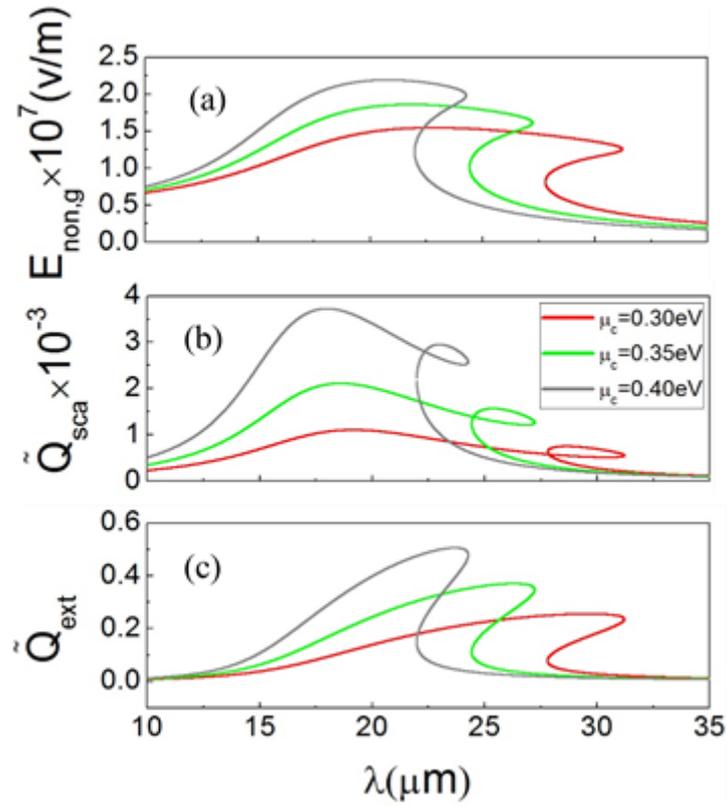

**Fig. 3. (a) Dependence of the modulus of the nonlinear field $E_{non,g}$ inside the graphene on the incident wavelength at different chemical potential; (b) and (c) illustrate the nonlinear far-field spectra versus incident wavelength at different chemical potential.**

Next, we also study the dependence of far-field properties of this graphene coated cylinder on the incident wavelength. Substituting the nonlinear coefficients $\tilde{A}_n$ and $\tilde{C}_n$ into Eq. (6) and Eq. (11), we have the nonlinear scattering and extinction efficiency in the FWST and QL [i.e. Eq. (15) and Eq. (16)]. Different from the near-field hysteresis spectra, bistable behavior for the nonlinear scattering efficiency is complex. As shown in Fig. 3 (b), the spectra is characterized by a "hysteresis loop", which indicates the bi-state do exists in the graphene wrapped cylinder. In addition, compared with Fig. 2 (c), it is found that for same parameters, the maximum value of linear scattering efficiency is almost 50 times larger than that of nonlinear scattering efficiency and the linear resonant wavelength is also larger than the nonlinear one, suggesting a different energy transfer mechanism when introduce the nonlinearity of graphene in this structure. Furthermore, the nonlinear extinction spectra is plotted in Fig. 3 (c), which exhibits similar bistable curves in Fig. 3 (a), and the values of the extinction efficiency is much larger than the scattering efficiency, showing a good absorption property of the graphene coated cylinder, which promise such structure a candidate for optical absorber.

Except the differences, both the peak value, the resonant wavelength and the unstable regions of the nonlinear scattering and extinction efficiency change in the same way as the nonlinear local field when the chemical potential of graphene changes, which reveals the filed-dependent property of the nonlinear scattering and extinction efficiency.

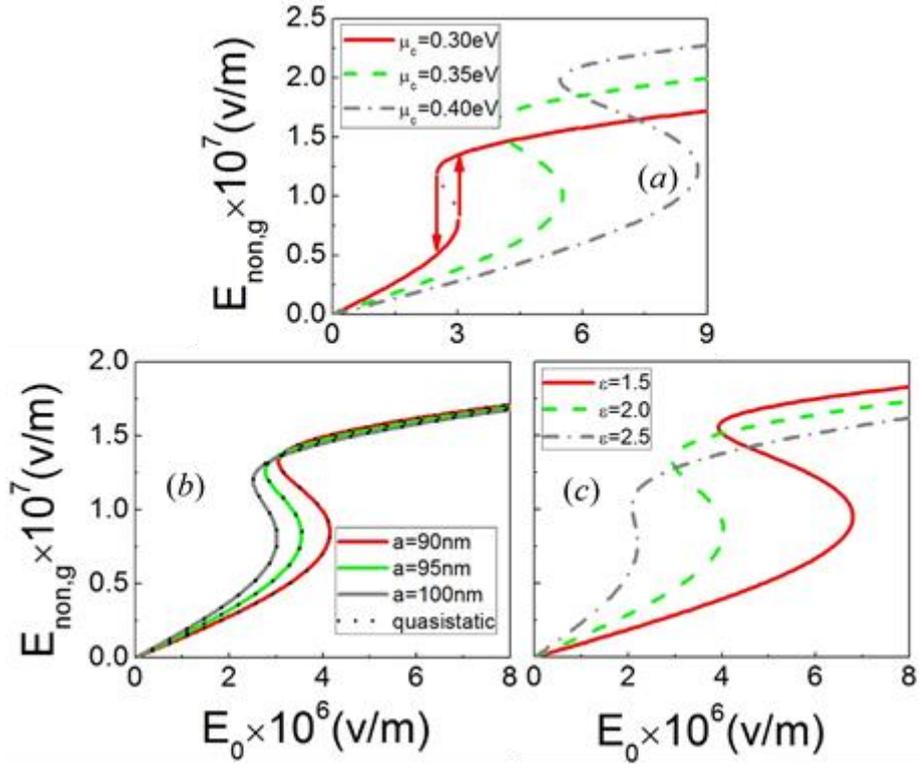

**Fig. 4.** The modulus of the nonlinear local field $E_{non,g}$ as a function of the external applied field $E_0$ for varied (a) chemical potentials; (b) sizes and (c) permittivities. Other parameters are $\varepsilon_h = 2.25$ and $\tau = 0.1\text{ps}$.

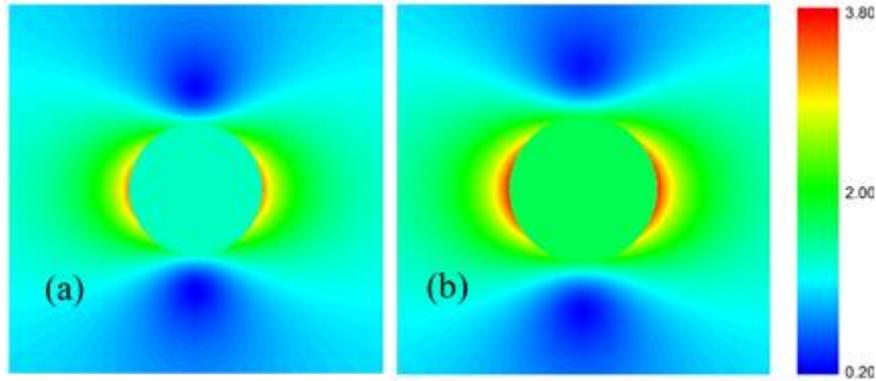

**Fig. 5.** Distributions of the electric fields inside and outside the nanocylinder for (a) $a = 90\text{nm}$, (b) $a = 100\text{nm}$. Other parameters are $\lambda = 25\mu m$, $\varepsilon = \varepsilon_h = 2.25$ and $\tau = 0.1\text{ps}$

    The intensity of the nonlinear local field in graphene depends not only on the incident wavelength, but also on the intensity of the incident electric field. Hence, we display the field-field patterns shown by Fig.4 with the incident wavelength fixed at $\lambda = 25\mu m$, which is near the resonant wavelength, where the nonlinear plasmonic resonance enhanced field would boost the nonlinear conductivity of graphene [see Eq. (13) and Eq. (14), where the third-order nonlinear term of the graphene conductivity $\sigma_3$ used to be much smaller than the linear term]. Fig. 4 (a) shows the dependence of the modulus of the nonlinear local field in graphene on the incident field. Take $\mu_c = 0.3\text{eV}$ for example, with the external increases to $3\times10^6 (v/m)$ [i.e. threshold up value] the nonlinear field jump discontinuously from $0.75\times10^7 (v/m)$ to $1.4\times10^7 (v/m)$, on the contrary, as one decrease the input field to $1.2\times10^6 (v/m)$ [i.e. threshold down

value], the nonlinear field decrease from $1.2 \times 10^7 (v/m)$ to $0.5 \times 10^7 (v/m)$ directly. Compared to the other two hysteresis curves in Fig. 5 (a), we find that the threshold value increases with increasing the chemical potential, which is because the third-order nonlinear term $\sigma_3$ gets small as the chemical potential adds, one wants to achieve the hysteresis spectra should increase the nonlinear local filed to make the term $\sigma_3 |E_{non,g}|^2$ comparable to the linear term, hence the incident filed increases. In addition, the unstable region becomes broader when enhance the chemical potential, showing an opposite behavior illustrated in Fig. 3 (a).

Since the nonlinear plasmonic resonance is also influenced by the structure properties of the cylinder, Fig. 4 (b) and (c) depict different bistable curves by tuning the sizes and relative dielectric constant of graphene. As shown in Fig. 4 (b), both the results under the FWST (solid line) and the QL (dotted line) are plotted, and show quite a good match, which means both the two theories are adoptable to investigate the bistable property of our structure within our parameter space. Besides, the threshold value decreases with the size increases, along with narrowed unstable region, which can be understood as increasing the particle size leads to enhancement of the linear field in the graphene as shown in Fig. 5, where the local field inside the graphene with cylinder size $a = 100 nm$ is larger than $a = 90 nm$, and with such basic filed enhancement, one would have a low incident filed to realize the bistability, hence a decreased threshold value. Fig. 4 (c) shows a same behavior with Fig. 4 (b) and, on the other hand, an opposite behavior to Fig. 4 (a), which suggests that enlarging the permittivity inside the cylinder equals to increase the cylinder size or decrease the chemical potential of graphene. In view of possible technological applications, this finding is expected to be very useful.

Finally, Fig. 6 (a) also shows the nonlinear local field inside the graphene as a function of the incident field within the FWST and the QL [i.e. still the Eq. (12) and Eq. (13)], as we can see, when the applied field intensity gets stronger [much more than $5 \times 10^7 (v/m)$ within our parameter space], there is another hysteresis curve based on the FWST, while the spectra from the QL illustrates that the nonlinear local field undergoes a monotonically increase. As a matter of fact, in small particle limit [29], the Bessel function and Hankel function can be expanded as a function of the dimensionless parameter $x$ and if we only consider the first several terms of the expansions, results derived from the FWST will turn to those from the QL. Meanwhile, our proposed parameters happen to meet such limit, hence the expression for nonlinear field [i.e. Eq. (13)] become more complicated, especially the coefficient $\tilde{F}_n$. Based on this, one can give the following explanation: as we increase the input electric field intensity, the contributions from other terms in the expansion function are negligible compared to the first several terms, hence results from the two theories are almost same, and this is what we see in Fig. 4 (b), on the contrary, when we further increase the external filed, the other terms play the leading role, hence the Fig. 6 (a). On the other hand, it can be proved in Fig. 6 (b) that the latter bistable curve changes in the same law as the former one when we vary the chemical potential of graphene, and one can expect that the latter spectra should be size- and permittivity- dependent. We think these results would offer a thorough understanding in realizing the optical bistability of the graphene wrapped dielectric cylinder.

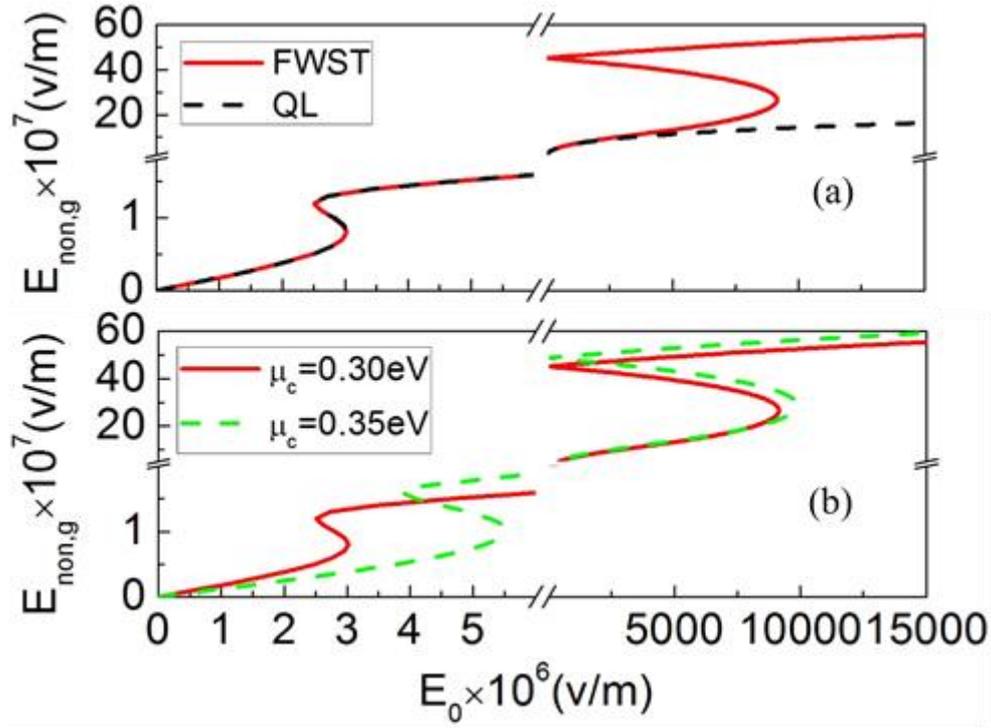

**Fig. 6.** The modulus of nonlinear local field $E_{non,g}$ as a function of the external applied field $E_0$, (a) under the FWST and QL with $a = 100\text{nm}$, $\mu_c = 0.3\text{eV}$; (b) with varied chemical potential, other parameters are $\varepsilon = \varepsilon_h = 2.25$ and $\tau = 0.1\text{ps}$.

## Conclusion

To conclude, we establish the nonlinear equations of near-field and far-field for nonlinear graphene wrapped dielectric cylinder in both the full wave theory and the quasistatic limit, and study the nonlinear optical bistable behaviors for the near-field, far-field scattering and extinction efficiency in such coated nanoparticle system. We find that introducing the nonlinearity of graphene would decrease the scattering efficiency and the resonant wavelength undergoes a blue shift. It's demonstrated that the two theories are both adoptable in analyzing the bistable behavior of the graphene coated structure when input a relevantly small electric field, however, once the field intensity is strong enough, results from the nonlinear full wave scattering theory turn to be more precise, which is very useful in practical applications. Moreover, it is shown that the threshold values are highly depend on the chemical potential of graphene besides size and dielectric constant of the particle, hence, it provides a new degree of freedom to control the local field and scattering (extinction) efficiency with the input one. All these novel properties have great potential for the design in optoelectronic switching and nano-memories.

**Acknowledgments**

This work was supported by the National Natural Science Foundation of China (Grant No. 11374223), the National Science of Jiangsu Province (Grant No. BK20161210), the Qing Lan project, "333" project (Grant No. BRA2015353), and PAPD of Jiangsu Higher Education Institutions.